\documentclass[letterpaper]{jpconf}
\pdfoutput=1
\usepackage{graphicx}
\usepackage{color}
\begin{document}
\title{Preliminary Results from Integrating Compton Photon Polarimetry in Hall A of Jefferson Lab}

\author{D Parno$^1$, M Friend$^1$, F Benmokhtar$^1$, G Franklin$^1$, R Michaels$^2$,\\ S Nanda$^2$, B Quinn$^1$ and P Souder$^3$}
%\author{}
\address{$^1$Carnegie Mellon University, Physics Department, 5000 Forbes Ave, Pittsburgh PA 15213, USA\newline
$^2$Thomas Jefferson National Accelerator Facility, 12000 Jefferson Ave, Newport News VA 23606, USA\newline
$^3$ Syracuse University, Department of Physics, Syracuse NY 13244, USA}

\ead{dparno@uw.edu}

\begin{abstract}
A wide range of nucleon and nuclear structure experiments in Jefferson Lab's Hall A require precise, continuous measurements of the polarization of the electron beam. In our Compton polarimeter, electrons are scattered off photons in a Fabry-Perot cavity; by measuring an asymmetry in the integrated signal of the scattered photons detected in a GSO crystal, we can make non-invasive, continuous measurements of the beam polarization. Our goal is to achieve 1\% statistical error within two hours of running. We discuss the design and commissioning of an upgrade to this apparatus, and report preliminary results for experiments conducted at beam energies from 3.5 to 5.9 GeV and photon rates from 5 to 100 kHz.

%We will discuss the design and analyzing power of our method and present preliminary polarization results for experiments conducted at beam energies from 1.05 to 5.9 GeV and photon rates from 5 to 100 kHz.
\end{abstract}

\section{Introduction}

The Thomas Jefferson National Accelerator Facility~\cite{Leemann:CEBAF01} houses a superconducting machine presently capable of delivering up to 200 $\mu$A of highly polarized electrons, accelerated to energies between approximately 1 and 6 GeV, to three experimental halls simultaneously. In each of these halls, the last few years have seen a wide variety of experiments requiring accurate knowledge of the electron beam polarization; in the largest hall, known as Hall A~\cite{Alcorn:HallA04}, the recent polarized-beam program has ranged from investigations of neutron spin structure~\cite{Zheng:d2nProposal} to the strange content of the nucleon~\cite{HAPPEXiiiProposal} to a measurement of the neutron distribution in heavy nuclei~\cite{PREXProposal}.

The longitudinal polarization of an electron beam may be defined as

\begin{equation}
P_e = \frac{N_e^+ - N_e^-}{N_e^+ + N_e^-}
\end{equation}

\noindent where $N_e^{+(-)}$ is the number of electrons with spin parallel (antiparallel) to the beam direction. In Hall A, the polarization of beams at GeV energies is measured via M{\o}ller scattering ($e^- e^- \rightarrow e^- e^-$) or via Compton scattering ($e^- \gamma \rightarrow e^- \gamma$), both of which are sensitive to the relative spins of the incident particles. For head-on Compton scattering between electrons with longitudinal polarization $P_e$ and photons with circular polarization $P_{\gamma}$, one may form an asymmetry between the energy-weighted, integrated Compton signal $S$ for time intervals in which the polarizations are parallel ($\uparrow \uparrow$) and antiparallel ($\uparrow \downarrow$):

\begin{equation}
A_{meas} = \frac{S^{\uparrow \uparrow} - S^{\uparrow \downarrow}}{S^{\uparrow \uparrow} + S^{\uparrow \downarrow}} = P_e P_{\gamma} \langle A_S \rangle
\label{basic_asy}
\end{equation}

\noindent where $\langle A_S \rangle$, the analyzing power, is the signal asymmetry that would be measured if the incident electron and photon beams were 100\% polarized~\cite{Passchier:NIKHEFCompton98}.

Compared to Hall A's M{\o}ller polarimeter~\cite{Glamazdin:Moller1999}, which uses a solid target, the major advantage of Compton polarimetry is its negligible effect on the electron beam as a whole; in our apparatus, approximately one electron in $10^9$ undergoes Compton scattering. Since the measurement is nondestructive, the beam polarization can be continuously monitored throughout the course of an experiment.

\section{Apparatus}

\begin{figure}[tb]
  \centering
    \includegraphics[width=0.7\textwidth]{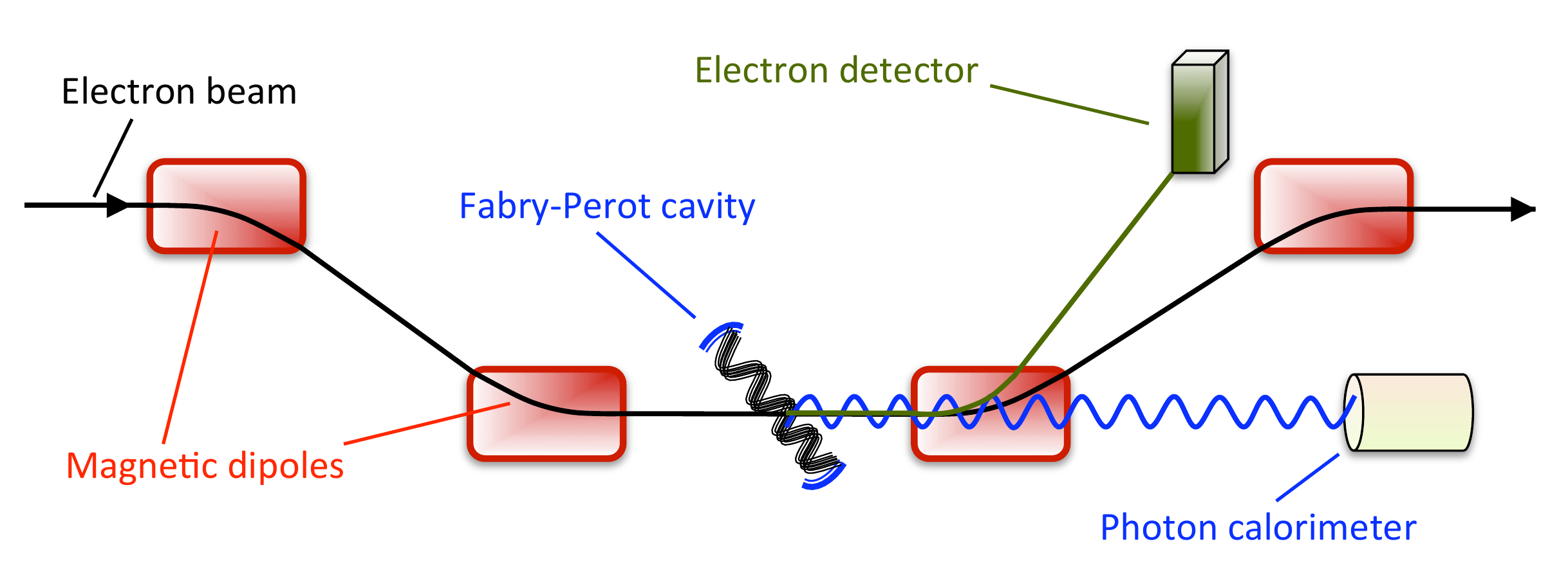}
    \caption{A schematic of the Compton polarimeter layout in Hall A. For clarity, angles have been exaggerated and distances are not to scale.}
  \label{fig:schematic}
\end{figure}

Our apparatus represents a partial upgrade of the Compton polarimeter installed in the Hall A beamline in 1999~\cite{Escoffier:ComptonEBP05}. Figure~\ref{fig:schematic} is a schematic representation of the polarimeter's layout. The beam enters from the left and is routed through a chicane formed by four magnetic dipoles. At the center of the chicane, electrons undergo Compton scattering with circularly polarized photons in resonance in a Fabry-P{\'e}rot cavity fed by an infrared laser ($\lambda = 1064$ nm)~\cite{Falletto:ComptonFPC01}; this system was upgraded to use green light ($\lambda = 532$ nm) in 2010. The crossing angle between the two beams is 23 mrad. The photon polarization is periodically flipped between right- and left-circular in order to control for systematic effects; during a flip, we characterize background processes by shutting off the Pound-Drever-Hall feedback loop~\cite{Drever:PDHTechnique83} between the cavity and the laser. This takes the cavity out of resonance, leaving negligible photon power at the Compton interaction point. Unscattered electrons, separated from the Compton-scattered particles by the third dipole in the chicane, continue on into the hall for the primary experiment. Scattered electrons may be measured in a silicon microstrip detector; in the remainder of this paper, we will discuss the detection and analysis of Compton-scattered photons measured in a calorimeter.

\begin{figure}[tb]
\begin{minipage}{0.45\textwidth}
\includegraphics[width=\textwidth]{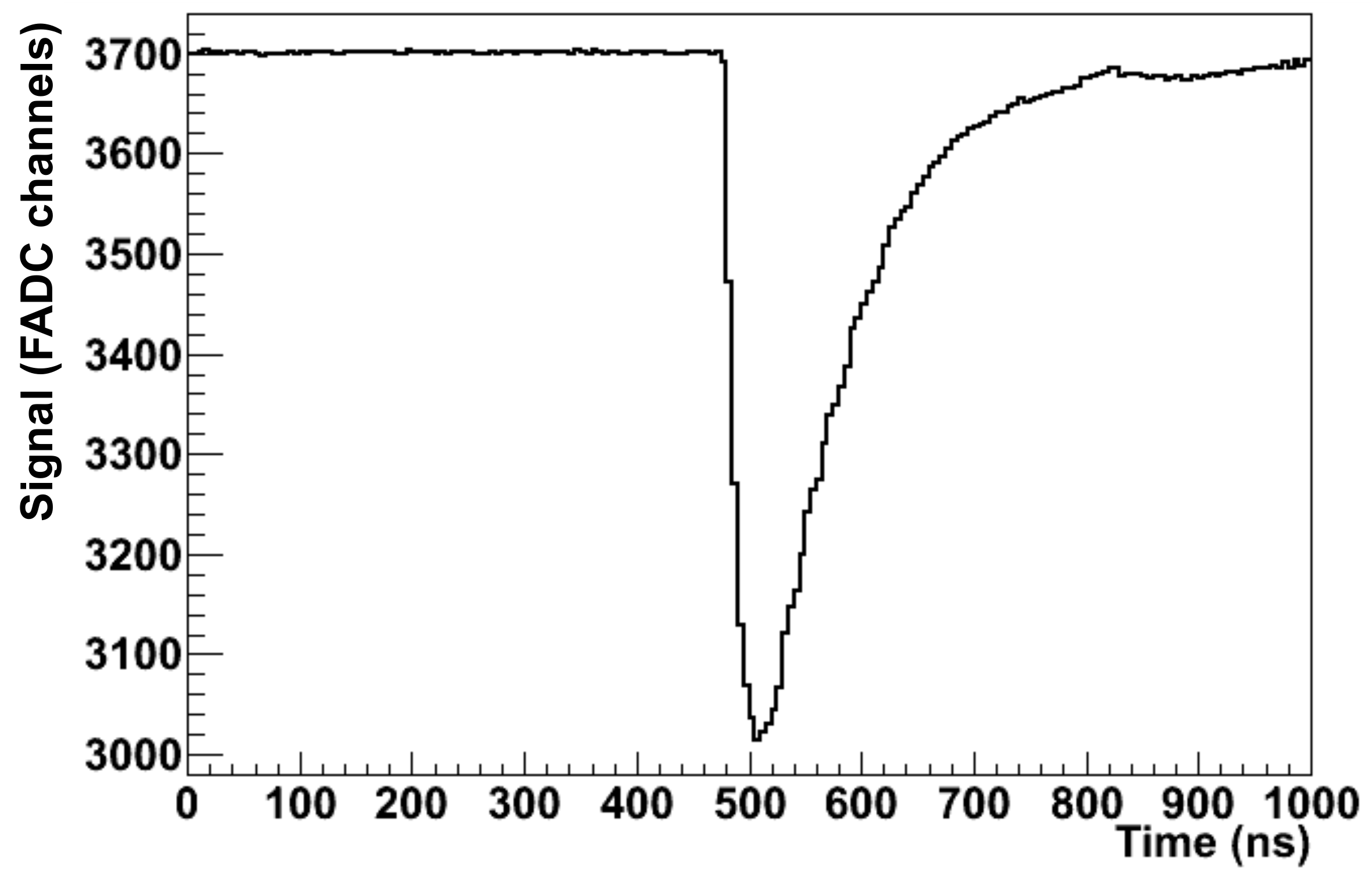}
\caption{\label{fig:pulse}A typical waveform from the GSO calorimeter for an incident photon in the energy range for Compton-scattered photons.}
\end{minipage}\hspace{0.1\textwidth}%
\begin{minipage}{0.45\textwidth}
\includegraphics[width=\textwidth]{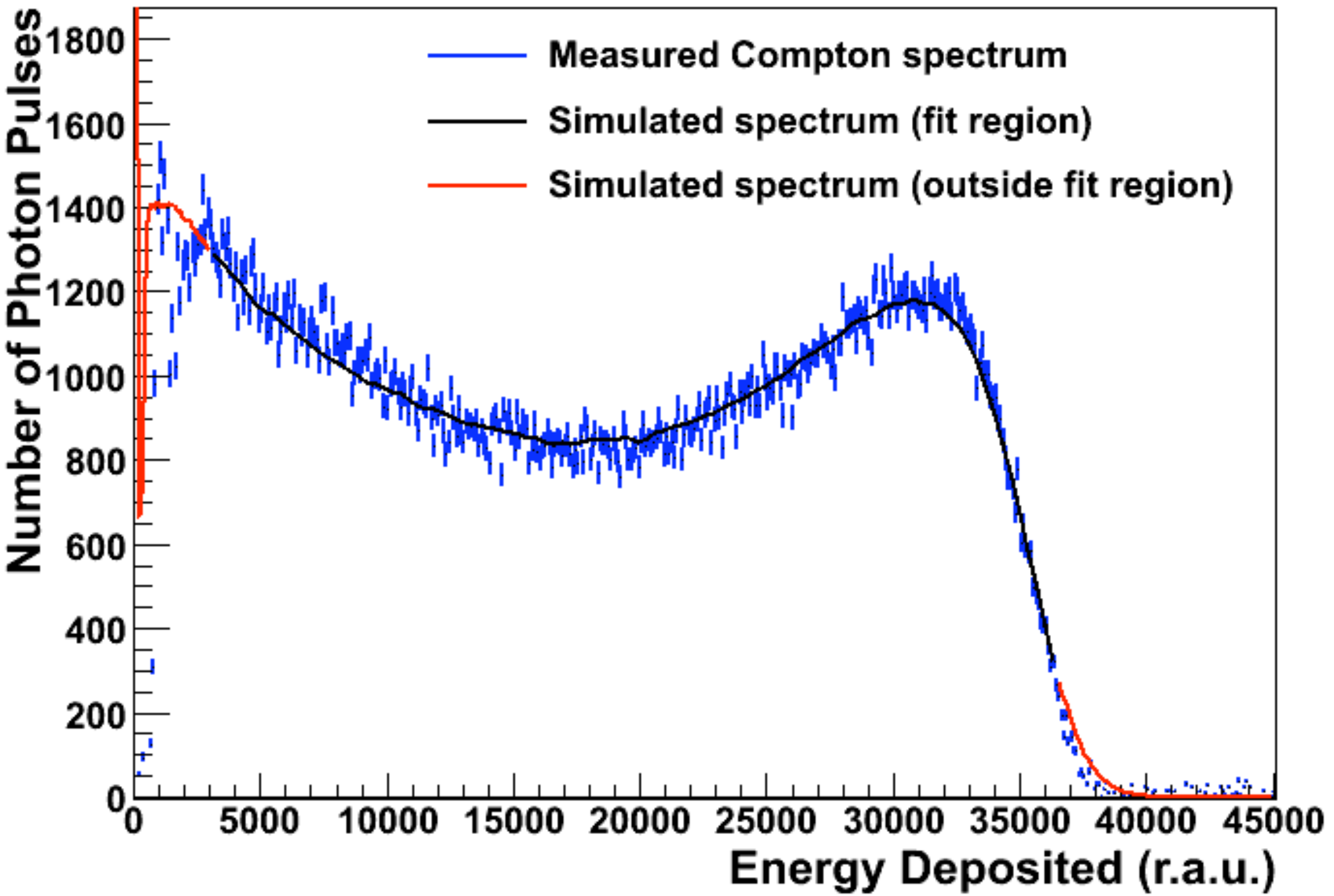}
\caption{\label{fig:spectrum}A measured, background-subtracted photon spectrum, compared to a GEANT4 simulation of a Compton-scattered photon spectrum. The raw ADC units of the x-axis are proportional to the photon energy.}
\end{minipage} 
\end{figure}

We have replaced the original Compton photon calorimeter, a 5x5 array of 2 cm x 2 cm x 23 cm PbWO$_4$ crystals~\cite{Neyret:ComptonCalorimeter00}, with a single cylinder of cerium-doped Gd$_2$SiO$_5$ (GSO) manufactured by Hitachi Chemical. With a 6-cm diameter and 15-cm length, this crystal is large enough to contain most of the shower from an incident photon, without the extended cross-calibration and gain matching required for a crystal array. The calorimeter is located approximately 6 m downstream of the Compton interaction point, and is mounted on a motorized table with remote-controllable motion along both axes (horizontal and vertical) transverse to the beam direction. Two narrow converter-scintillator pairs allow precise centering on the beam of Compton-scattered photons, which forms a cone with higher-energy photons at the center. Figure~\ref{fig:pulse} shows the shape of a typical pulse from the GSO's PMT for a photon in the energy range of interest. Figure~\ref{fig:spectrum} shows the distinctive Compton-scattered photon energy spectrum measured in the GSO for an electron beam energy of 5.7 GeV. Also plotted is the spectrum resulting from a GEANT4 simulation of the Compton apparatus, fit to the experimental spectrum with three parameters: amplitude, gain, and detector resolution. The sharp drop at high scattered-photon energies corresponds to the Compton edge, the maximum energy $k'_{max}$ that can be carried by a Compton-scattered photon. 

\section{Integrating Data Acquisition}

\begin{figure}[tb]
\begin{minipage}{0.45\textwidth}
    \includegraphics[width=\textwidth]{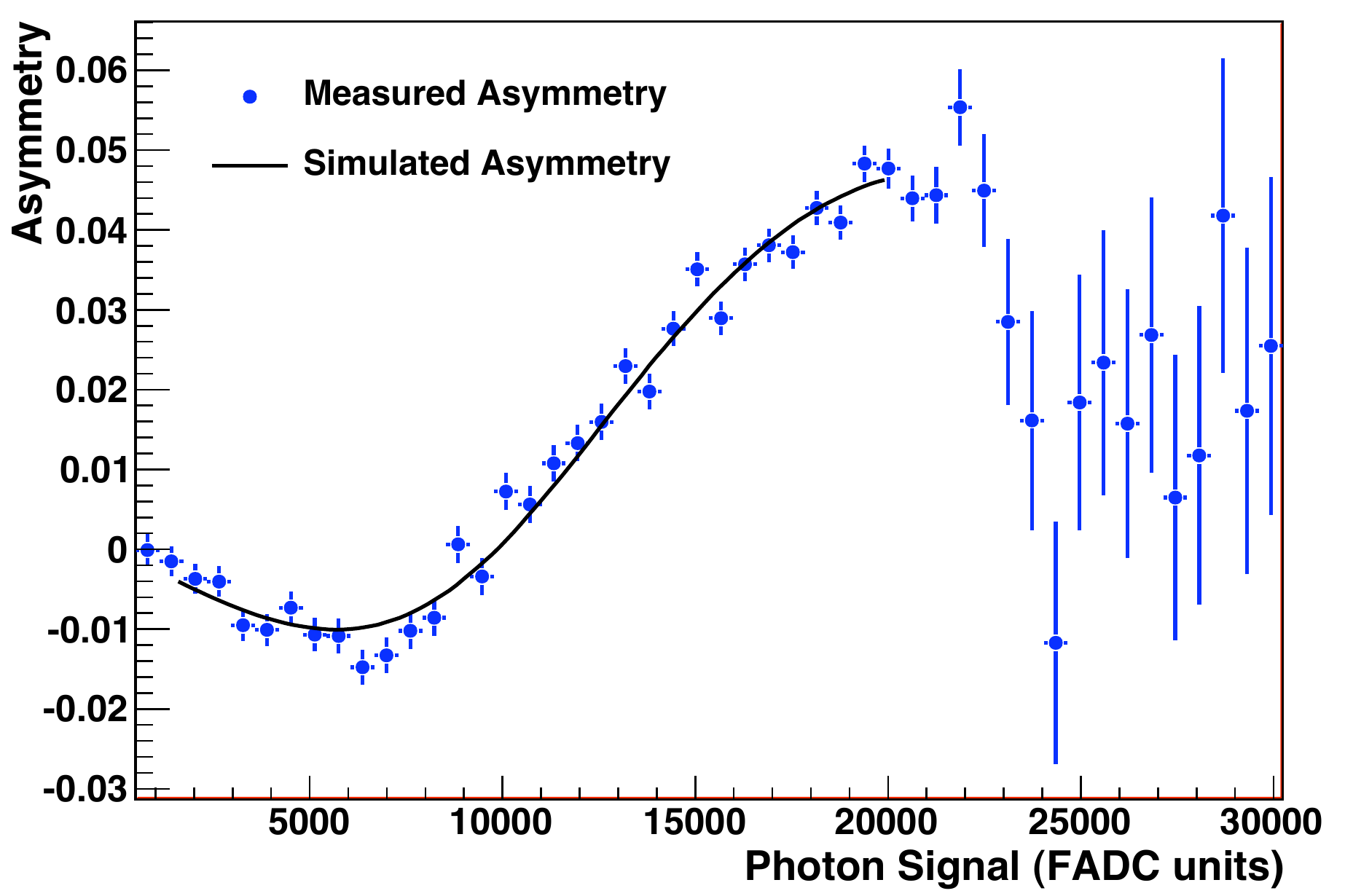}
    \caption{\label{fig:analyzingpower}The simulated theoretical asymmetry $A_S$ as a function of scattered photon energy $k'$, compared to the observed counting asymmetry in successive photon energy bins. $A_S$ rapidly approaches zero as the observed photon energy exceeds $k'_{max}$.}
  \end{minipage}\hspace{0.1\textwidth}%
\begin{minipage}{0.45\textwidth}
\includegraphics[width=\textwidth]{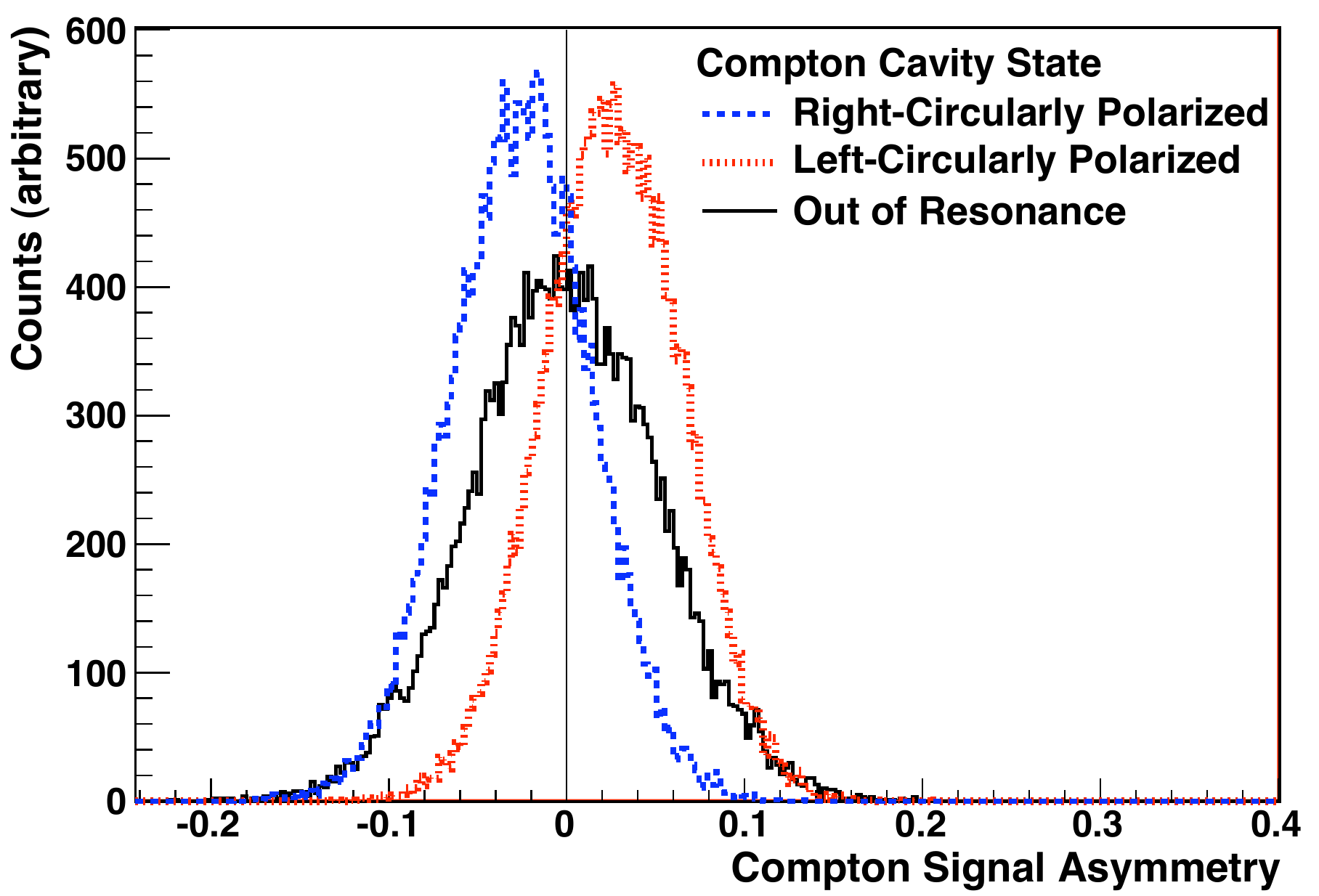}
\caption{\label{fig:asyRL}Integrated, energy-weighted Compton asymmetries computed over electron beam helicity pairs (each 67 ms in duration) over the course of a two-hour run. The sign flip between the asymmetries for the two photon polarization directions is based on the conventional definition of the asymmetry.}
\end{minipage} 
\end{figure}

As a function of scattered photon energy $\rho = k'/k'_{max}$, the theoretical asymmetry $A_S$ is given by

\begin{equation}
A_S = \frac{2\pi r_0^2 a}{d\sigma / d\rho} \left( 1 - \rho \left( 1 + a \right) \right) 
	\left[1 - \frac{1}{\left( 1 - \rho\left( 1 - a \right) \right)^2} \right]
\end{equation}  

\noindent where $a = \left( 1 + 4kE_e / m_e^2 \right) ^{-1}$, $r_0$ is the classical electron radius, $m_e$ is the electron mass, and $k$ and $E_e$ are the incident energies of the photon and electron, respectively~\cite{Denner:ComptonRadiative99,Bardin:ComptonCDR96}. (Although this equation is derived for head-on scattering, the small angle between photon and electron beams in our apparatus has a negligible effect.) Figure~\ref{fig:analyzingpower} shows the evolution of $A_S$ with photon energy. For low values of $\rho$, $A_S$ is slightly negative, before crossing zero and reaching a maximum at $k' = k'_{max}$ or $\rho = 1$.

The shape of the $A_S$ curve suggests that it is advantageous to compute an asymmetry in the energy-weighted integral of the photon signal, rather than in the raw counting rates. In such an integral, the greatest contributions to the asymmetry will be made by photons in the energy range where the analyzing power is at a maximum. Furthermore, because this integral is less sensitive to low-energy uncertainties in the detector response function, the method is well-suited to stand-alone photon detector running: accurate asymmetries may be measured even without calibration against the scattered-electron detector.  We therefore designed and commissioned a data acquisition system (DAQ) with both integrating and counting capability to replace the counting DAQ of the original Hall A Compton photon system.

The new DAQ is based on a 12-bit FADC from Struck DE. A photon pulse registers as a negative waveform relative to a programmable pedestal level, as in Figure~\ref{fig:pulse}. The signal from the photon detector is sampled at 200 MHz and integrated in a digital summing accumulator for a single helicity window, an interval in which the electron beam helicity is well-defined; the helicity flips at fixed rates that range from 30 Hz to 1 kHz, depending on experimental requirements. (The results shown here are all from 30 Hz running.) We then form an asymmetry in the pedestal-subtracted, summed signal between intervals where the incident particles have opposite polarization configurations, according to Equation~\ref{basic_asy}; this energy-weighted integral is proportional to the degree of longitudinal polarization of the electron beam.

%The total signal from an accumulator at the end of a helicity window is given by

%\begin{equation}
%Acc = N \left( \bar{P} - \bar{S} \right)
%\end{equation}

%\noindent where $N$ is the number of 5-ns samples that contributed to the integral, $\bar{P}$ is the average pedestal value, and $\bar{S}$ is the average physics signal; the total signal for an interval can be extracted as $N \bar{S} = N \bar{P} - Acc$. An asymmetry in $N \bar{S}$ is then formed between intervals where the incident particles have opposite polarization configurations, according to Equation~\ref{basic_asy}.  

In addition to an accumulator that sums all signal over an interval, auxiliary hardware accumulators allow us to examine only contributions from particular types of signal, such as very small signals (e.g. pedestal noise) or very large ones. These regions are delineated by setting appropriate values for two programmable thresholds. We may also program the FADC to store a certain number of samples in memory; for example, when the signal crosses a threshold, the $N_{\mathit{before}}$ samples prior to the threshold crossing (and the $N_{\mathit{after}}$ samples subsequent to crossing the threshold in the other direction) can also be added to the integral, allowing integration over the entire pulse rather than its tip alone. Altogether, the various combinations of thresholds and timing data yield five auxiliary accumulators, each integrating over a particular subset of the signal.

%Altogether, five additional hardware accumulators using different combinations of thresholds and timing data are available.

We are also able to store a small number of digitized, 500-ns waveforms per interval, both from prescaled triggers (Figure~\ref{fig:pulse}) and from random intervals, as well as the numerical integrals of up to 300 prescaled photon pulses per integral, depending on the rate of helicity flips. This capability allows us to perform analyses on individual photon pulse data (such as the Compton spectrum plotted in Figure~\ref{fig:spectrum} and the energy-binned counting asymmetry plotted in Figure~\ref{fig:analyzingpower}) in addition to the primary, integrating analysis. For example, the counting asymmetry may be fit to a Monte Carlo asymmetry  to make an independent measurement of the beam polarization.

\section{Analysis and Preliminary Results}

The asymmetry defined in Equation~\ref{basic_asy} may be formed between any two intervals with opposite spin configurations for the incident particles. The shortest timescale on which we can form an asymmetry is between two adjacent electron helicity windows; depending on the helicity flip rate, a complete helicity pair may comprise between 8.3 and 67 ms. By adding the contributions from helicity windows with the same spin configurations, however, we may also form asymmetries on longer timescales: over the course of a two-minute interval with a single photon polarization direction, for example, or over a several-hour run. The choice of asymmetry timescale will affect the apparent stability of asymmetries from interval to interval.

Figure~\ref{fig:asyRL} shows the distribution of energy-weighted integrated asymmetries, computed over helicity pairs of 67 ms each, for two laser polarization states (right-circular and left-circular) as well as a background measurement taken with the cavity out of resonance. The magnitudes of the two independent asymmetry measurements can be averaged to find the asymmetry for the entire two-hour run. Data taken during beam trips, high-voltage trips, and polarization transitions have been excluded. Meanwhile, the measurement of the background signal $B$ allows us to correct the Compton signal by accounting for the dilution of the asymmetry due to background processes (primarily bremsstrahlung and synchrotron radiation):

\begin{equation}
A_{meas} = \frac{\left( S^{\uparrow \uparrow} - B^{\uparrow \uparrow} \right) - \left( S^{\uparrow \downarrow} - B^{\uparrow \downarrow} \right)}{\left( S^{\uparrow \uparrow} - B^{\uparrow \uparrow} \right) + \left( S^{\uparrow \downarrow} - B^{\uparrow \downarrow} \right)} = P_e P_{\gamma} \langle A_S \rangle
\label{corr_asy}
\end{equation}

\noindent Ideally, $B^{\uparrow \uparrow} = B^{\uparrow \downarrow}$, but in practice there may be so-called false asymmetries due to helicity-dependent changes in beam tune. 

We may find $\langle A_S \rangle$ by using GEANT4 to simulate the analyzing power as a function of scattered photon signal (as in Figure~\ref{fig:analyzingpower}) for our kinematics and apparatus, and performing an energy-weighted integral of the function. Meanwhile, $P_{\gamma}$ is measured from the light transmitted through the cavity when it is in resonance: two powermeters are placed at the outputs of a polarizing beam splitter in the optical path, and their relative readings are combined with the cavity transfer function to compute the photon polarization at the Compton interaction point.

\begin{figure}[tb]
\begin{minipage}{0.45\textwidth}
    \includegraphics[width=\textwidth]{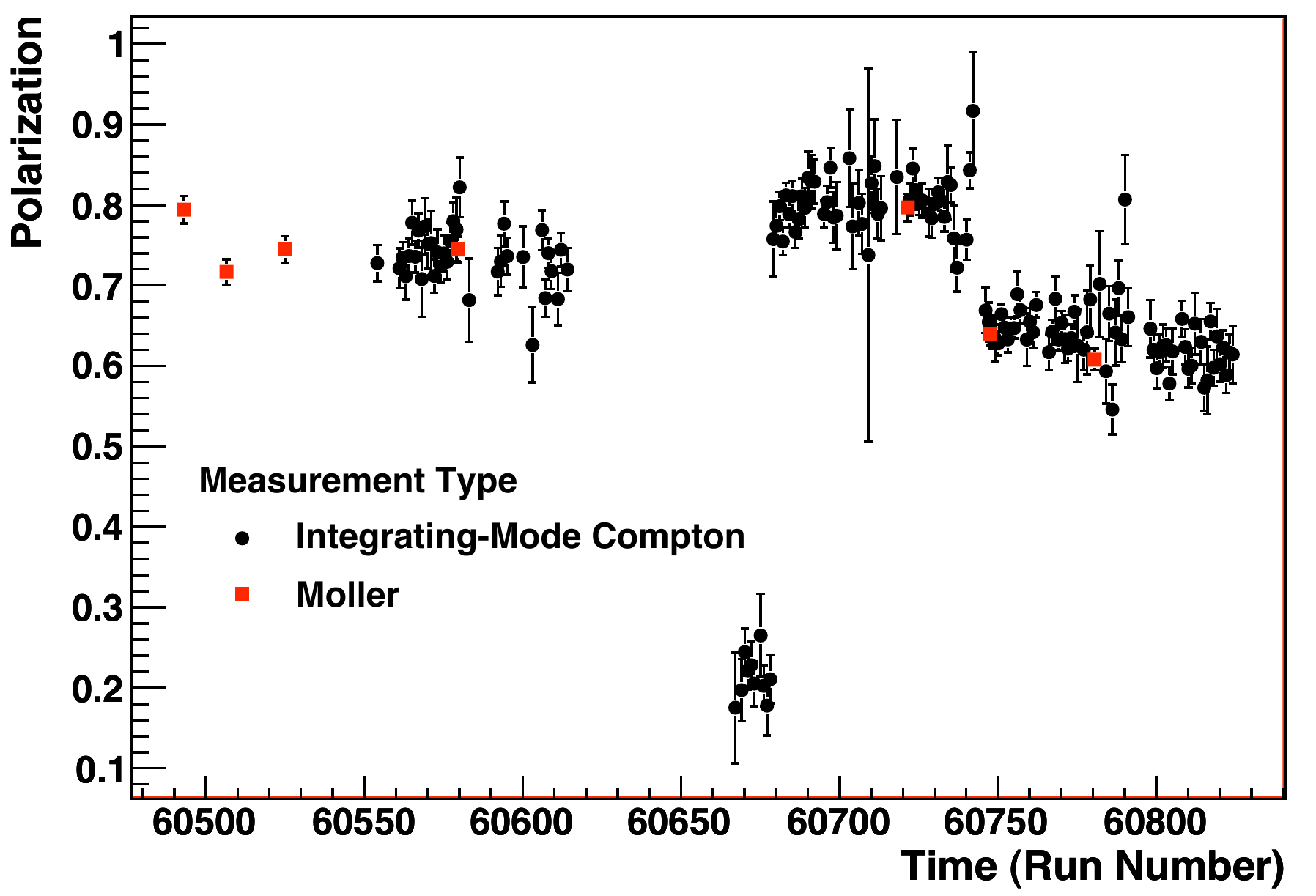}
    \caption{\label{fig:d2nhist}Preliminary electron beam polarization results from integrated Compton photon data during E06-014~\cite{Zheng:d2nProposal}, compared to measurements from the M{\o}ller polarimeter; the experiment ran for six weeks in February and March 2009. Error bars are statistical only. Asymmetries were computed over helicity pairs.}
\end{minipage}\hspace{0.1\textwidth}%
\begin{minipage}{0.45\textwidth}
	\includegraphics[width=\textwidth]{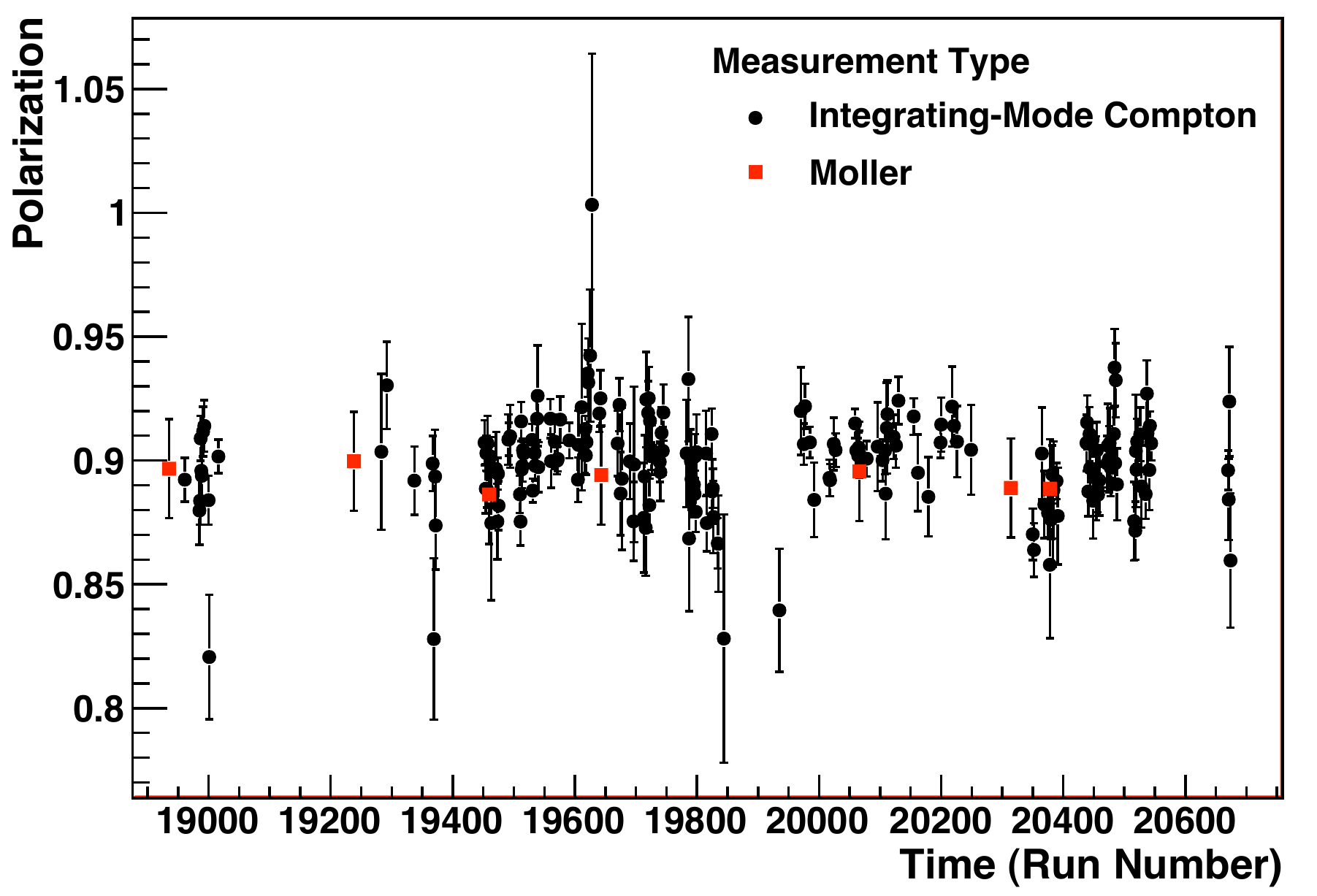}
	\caption{\label{fig:happexhist}Preliminary electron beam polarization results from integrated Compton photon data during HAPPEX-III~\cite{HAPPEXiiiProposal},  compared to measurements from the M{\o}ller polarimeter; the experiment ran for two months in September and October 2009. Error bars are statistical only. Asymmetries were computed over two-minute laser polarization states.}
\end{minipage}\hspace{0.1\textwidth}%sep 4 to 
\end{figure}

Figure~\ref{fig:d2nhist} shows preliminary beam polarization measurements from this method for an experiment in spring 2009. Our measurements agree well with those performed by the M{\o}ller polarimeter; the four configurations of the experiment, including a few days of nearly unpolarized beam, may be clearly distinguished. Asymmetries were computed over helicity pairs. At a production electron beam current of 15 $\mu$A, the detector saw average photon rates of 3 kHz of background (with the cavity out of resonance) and 5 kHz with the cavity in resonance, giving relative statistical errors of about 3.6\% in two hours of running. We see similar success in Figure~\ref{fig:happexhist}, which shows preliminary polarization measurements for asymmetries computed over two-minute cavity cycles in fall 2009; this experiment ran with a beam current of 100 $\mu$A, yielding photon rates of about 55 kHz of background and 100 kHz with the cavity in resonance. The resulting statistical error was about 0.7\% in two hours of running.

\section{Conclusion and Outlook}

The integrating method is a powerful tool for Compton photon polarimetry, particularly at high rates and low electron energies. We have reported on the design and commissioning of a new integrating data acquisition and analysis system in Hall A of Jefferson Lab; our preliminary results indicate good agreement with measurements performed by existing polarimeters. As we work to quantify sources of systematic errors and refine our technique, we expect that the new photon detector, DAQ, and analysis software will continue to contribute to the polarized-beam physics program in Hall A.

\ack
This work was supported by DOE grants DE-AC05-06OR23177, under which Jefferson Science Associates, LLC, operate Jefferson Lab, and DE-FG02-87ER40315.

\section*{References}
\bibliography{2010_INPC_bib}

\end{document}